      [1999/12/01 v1.4c Il Nuovo Cimento]
\documentclass{cimento}
\usepackage{amsmath} 
\usepackage{amsfonts} 
\usepackage{amssymb} 

\title{Electro-weak Model within the framework of\\ Lorentz gauge theory: Ashtekar variables?}
\author{O.M. Lecian\from{ins:x}\from{ins:y}\thanks 
{lecian@icra.it} and G. Montani\from{ins:x}\from{ins:y}\from{ins:z} 
\thanks{montani@icra.it}} 
\instlist{\inst{ins:x} Universit\`a di Roma ``La Sapienza'', Dipartimento di Fisica \\
Piazzale A.Moro 5-00185 Roma, Italy
  \inst{ins:y} ICRA---International Center for Relativistic Astrophysics\\
  c/o Universit\`a di Roma ``La Sapienza'', Dipartimento di Fisica  
  \inst{ins:z} ENEA C.R. Frascati (Dipartimento F.P.N.),\\ Via Enrico Fermi 45-00044 
Frascati, Roma, Italy }
\PACSes{\PACSit{11.15.-q}  {04.50.+h}}
\begin{document}

\maketitle

\begin{abstract}
The Electroweak (EW) model is geometrized in the framework of a 5D gauge theory of the Lorentz group, after the implementation of the Kaluza-Klein (KK) paradigm. The possibility of introducing Ashtekar variables on a 5D KK manifold is considered on the ground of its geometrical structure.

\end{abstract}
\section{Introduction}
The introduction of a gauge theory of the Lorentz group is based on the different behaviour of spinors and tensors under local Lorentz transformation, even in flat space-time. If this gauge theory is implemented in a 5D KK scenario, where spinors can be illustrated to have chiral features, extraD generators of the Lorentz group have to be defined, in order to fulfill the proper commutation relations: this task is achieved establishing new fundamental 5D spinor doublets. As a result, the EW model finds a geometrical description.\\
Two ADM splittings performed on the KK manifold and the study of the 5D Lorentz group will provide the basis for the definition of Ashtekar variables in this framework, through the search of evolutionary quantities.  
\section{5D KK theories}
Let $V^{5}=V^{4}\oplus S^{1}$ be a 5D KK manifold, and let
\small
\begin{equation}\label{basse}
\begin{cases}
j_{55}=1\cr
j_{5\mu}=g'k\tilde{B}_{\mu}\cr
j_{\mu\nu}-(\tilde{g}'k)^{2}\tilde{B}_{\mu}\tilde{B}_{\nu}=g_{\mu\nu},\cr
\end{cases}
\end{equation}
\normalsize
be the 5D metric tensor, where $k$ and $\tilde{g}'$ are constants. The behavior of $\tilde{B}_{\mu}$ can be investigated by considering the set of allowed coordinate transformations 
\begin{equation}\label{ammesse}
x'^{\mu}=x'^{\mu}(x^{\rho});x'^{5}=x^{5}+\alpha (x^{\rho})
\end{equation}
that constrains $j_{55}$ to be a scalar (which we set equal to one) and verifies the cylindrical hypothesis: under these assumptions, the field $\tilde{B}_{\mu}$ behaves as a 4D vector under pure 4D coordinate transformations, while, for extra-D coordinate transformations, it acts as a gauge field, provided that $\alpha(x^{\rho})=\tilde{g}'k\alpha'(x^{\rho})$, so that $\tilde{B}_{\mu}\rightarrow \tilde{B}_{\mu}+\partial_{\mu}\alpha'(x^{\sigma})$. This is how an Abelian U(1) interaction is geomterized, and $\tilde{g}'$ can be interpreted as its coupling constant.\\
Tetradic vectors\footnote{Throughout sections 2-5, we have adopted the following notation:\\
-lower-case Greek letters for 4-dimensional world indices, $\mu=0,1,2,3$;\\
-upper-case Greek letters for 5-dimensional world indices, $\Omega=0,1,2,3,5$;\\
-overbarred lower-case Latin letters for 4-dimensional bein indices, $\bar{a}=0,1,2,3$;\\
-overbarred upper-case Latin letters for 5-dimensional bein indices, $\bar{A}=0,1,2,3,5$;\\
- $i,j,k$ for $1,2,3$;\\ 
- natural units, $\hbar \equiv c \equiv 1$, exept for when dimensional analysis is needed.} $\left\{V^{\bar{A}}_{\Omega}\right\}$ obey the request $j_{\Omega\Pi}=V^{\bar{A}}_{\Omega}V^{\bar{B}}_{\Pi}\eta_{\bar{A}\bar{B}}$, and read
\small
\begin{equation}
\begin{cases}
V_{5}^{\bar{a}}=0\cr V_{5}^{\bar{5}}=1\cr V^{\bar{5}}_{\mu}=\tilde{g}'k \tilde{B}_{\mu}\cr V_{\mu}^{\bar{a}} : g_{\mu\nu}=V_{\mu}^{\bar{a}}V_{\nu}^{\bar{b}}\eta_{\bar{a}\bar{b}}.\cr
\end{cases}
\end{equation}
\normalsize
In the following, the hyper-charge interaction will be geometrized in such a setting \cite{lem2006}.
\section{5D spinors}
The Lagrangian density of a massless fermion field can be extended to 5D as
\begin{equation}\label{13}
{^{4}L}=-\frac{i}{2}\bar{\psi_{i}}\gamma^{\mu}\partial_{\mu}\psi_{i} + H.C. \rightarrow
{^{5}L}=-\frac{i}{2}\bar{\chi_{i}}\gamma^{\Omega}\partial_{\Omega}\chi_{i} +H.C.=-\frac{i}{2}\bar{\chi_{i}}\gamma^{\bar{A}}\partial_{\bar{A}}\chi_{i}+H.C.:
\end{equation}
spinor fields can be specified to 5D as
\begin{equation}
{^{5}\chi}_{i}(x^{\rho},x^{5})=\frac{1}{\sqrt{L}}{^{4}\psi}_{i}(x^{\rho})e^{\frac{i2\pi N_{i}x^{5}}{L}},
\end{equation}
where the periodic dependence on the extra-ring fits the structure of $V^{5}=V^{4}\oplus S^{1}$; the Dirac algebra can be generalized to 5D by the matrix $\gamma^{5}=i\gamma^{0}\gamma^{1}\gamma^{2}\gamma^{3}$, which still satisfies the required anti-commutation relations $[\gamma^{\bar{A}},\gamma^{\bar{B}}]_{+}=2I\eta^{\bar{A}\bar{B}}$. If the 4D identity $I=P_{R} + P_{L}$, $P_{R/L}\equiv \frac{I\pm\gamma^{5}}{2}$ is inserted in the Lagrangian density above, two different Dirac equations are found for right- and left-handed spinor fields, ${^{5}\chi}_{iR}$ and ${^{5}\chi}_{iL}$, respectively:
the natural right/left asymmetry shown in 5D perfectly fits the chiral features of the hyper-charge chirality , thus suggesting us to interpret the dimensionless parameter $N_{i}$ as the different periodicity properties of the fields along the extra-ring.
\section{Lorentz gauge theories}
The need to introduce a Lorentz gauge field \cite{lmm2006} in curved space-time arises from the fact that, while spin connections $\Gamma_{\mu}^{R}$ are intended to restore the properties of Dirac matrices in the physical space-time, gauge connections $\Gamma_{\mu}^{K}$ allow one to recover invariance under local Lorentz transformations for spinor fields on the tangent bundle, thus implementing the covariant derivatives
\small
\begin{equation}
D_{\mu}\psi=\partial_{\mu}\psi-\Gamma_{\mu}\psi,\quad D_{\mu}\bar{\psi}=\partial_{\mu}\bar{\psi}+\bar{\psi}\Gamma^{\mu},
\end{equation}
\normalsize
where $\Gamma_{\mu}=\Gamma_{\mu}^{R}+\Gamma_{\mu}^{K}$. While spin connections are not gauge fields, gauge connection can be identified with suitable bein projections of the contortion field, $\Gamma_{\mu}^{K}=\frac{1}{2}A^{\bar{a}\bar{b}}_{\mu}\Sigma_{\bar{a}\bar{b}}$; appropriate Lagrangian densities must be introduced, which refer to the gravitational and to the non-Abelian fields, respectively.\\
When this formalism is extended to a 5D KK model, where the 5D Lorentz symmetry is broken, generators must be found, such that the structure of the manifold $V^{5}=V^{4}\oplus S^{1}$ is taken into account. Furthermore, 4D and extra-D generators, $\Sigma^{\bar{a}\bar{b}}$ and $\Sigma^{\bar{a}\bar{5}}$, respectively, must commute when applied to ${^{5}\chi}$: the extra-D generators must be proportional to the identity $I$; the EW model can be therefore geometrized by the introduction of new fundamental spinor fields, the doublets ${^{5}X}$, on which the action of the extra-D gauge fields is that of the non-Abelian SU(2) isospin gauge fields. Because, for each family of leptons $f_{i}$, left-handed doublets only,
\small
\begin{equation}\label{doppiettichilep}
X_{f_{i}L} \equiv
\left(
\begin{array}{c}
\chi_{\nu_{i}L}\\
\chi_{l_{i}L}
\end{array}
\right)\equiv \frac{1}{\sqrt{L}}\Psi_{iL}e^{\frac{i2\pi}{L}N_{iL}x^{5}},
\end{equation}
\normalsize
are requested in the EW model, the extra-D generators must contain the left-handed projectors $P_{L}$, thus casting right-handed doublets ${^{5}X}{R}$ to the null space of the extra-D generators, i.e. the ${^{5}X}_{R}$'s are unphysical objects. As a consequence, the extra-D generators read $\Sigma^{\bar{i}\bar{5}}\equiv ig\frac{\sigma^{i}}{2}P_{L}$, the extra-D bein projection of the contortion field $A^{\bar{i}\bar{5}}_{\mu}$ being identified with the weak-isospin gauge fields $W^{i}_{\mu}$, while $\Sigma^{\bar{0}\bar{5}}\equiv 0$, for the U(1) hyper-charge gauge field has already been geometrized via the standard KK mechanism. This particular choice of the generators doesn't violate any of the previous hypotheses.\\
The fields $A^{\bar{A}\bar{B}}_{5}$ must be vanishing quantities, in order not to have "Lorentz scalars" in the model. Thanks to this choice, the remaining fields meet the proper world transformation, while proper gauge transformations are obtained because of the vanishing commutators established above: the index $\bar{5}$ is therefore constrained to have a passive role.
\section{Dimensional reduction}
The total 5D KK action $^{5}S$ splits up into the ordinary Einstein action and the action of an Abelian gauge field
\small 
\begin{equation}\label{azionefinalmentedue}
S=\frac{-c^{3}}{16 \pi  {^{5}G}}\int dx^{5}d^{4}x\sqrt{-j}{^{5}R}=
\frac{-c^{3}}{16 \pi  {^{5}G}}\int d^{4}x\sqrt{-g}[^{4}R+(\frac{\tilde{g}'k}{2})^{2}\tilde{F}_{\mu\nu}\tilde{F}^{\mu\nu}],
\end{equation}
\normalsize
where $\tilde{F}_{\mu\nu} \equiv \partial_{\nu}\tilde{B}_{\mu}-\partial_{\mu}\tilde{B}_{\nu}$ is the bosonic Lagrangian density of the field $\tilde{B}_{\mu}$, and ${^{5}G}$ is related to the 4D Newton constant by $G=\frac{{^{5}G}}{L}$. Coupling with matter is reproduced via the bein projection of the ordinary derivative, $\partial_{\bar{\mu}}=e_{\bar{\mu}}^{B}\partial_{B}={^{4}\partial_{\bar{\mu}}}-g'kB_{\bar{\mu}}\partial_{5}$, thus discovering the relations between geometrical and gauge objects: $g'\equiv \tilde{g}'\frac{2\pi k}{L}$, $B_{\mu}\equiv \frac{\tilde{B}_{\mu}}{M}$, provided that $MN_{i}\equiv -y_{i}$, the hyper-charge of each field; if $N_{i}\in Z \forall i$, without loss of generalities, we can set $M\equiv\frac{1}{6}$, so that the periodicity is an integer sub-multiple of the circumference $L$. Extra-D derivatives vanish when coupled to $\gamma^{\bar{5}}$, i.e. $
\bar{\chi}_{a}\gamma^{\bar{5}}\partial_{\bar{5}}\chi_{a}\equiv 0$.\\
On the other hand, the bein projections of the covariant derivative $D_{\bar{A}}=e_{\bar{A}}^{\Omega}D_{\Omega}$ leads to the bein projection of the spin connections, $\Gamma^{R}_{\bar{a}}\equiv {^{4}\Gamma}_{\bar{a}}$,
$\Gamma^{R}_{\bar{5}}\equiv 0$.\\
Collecting all the terms together, and integrating in the extra-coordinate, the EW model and the 4D Lorentz gauge interaction in curved space-time are restored.\\
4D Lorentz transformations and $SU(2)\otimes U(1)$ gauge transformations are established for spinors by \ref{ammesse}, and conserved charges are then found according to the geometrization procedure of the gauge interaction. 
\section{Ashtekar formalism}
The possibility of extending the Ashtekar formalism \cite{a86},\cite{a87} to a 5D KK scenario, and then adopting it for the description of the 5D Lorentz group can be investigated if attention is payed to the geometry involved in this scheme. In order to state the most general formulation, all the metric components of the KK theory will be taken into account, i.e. the scalar field $g_{55}=\phi$ will not be considered constant any more, as well as the bein vector $e^{\bar{5}}_{5}$.\\
Evolutionary variables can be identified considering the Gauss-Codacci formula, which, in 5D, reads\footnote{Upper-case Latin letters denote space-like world indces, i.e. $I=1,2,3,5$, while
overbarred upper-case Latin letters are used for space-like bein indeces, $\bar{A}=1,2,3,5$;\\}
\begin{equation}
{^{5}R}={^{4}R}+K^{2}-K_{IJ}K^{IJ}+S.T.,\ \ I=1,2,3,5
\end{equation}
where $K_{IJ}$ is the extrinsic curvature,  
\begin{equation}
K_{IJ}=\frac{1}{2N}\left(D_{I}N_{J}+D_{J}N_{I}+\partial_{0}h_{IJ}\right)=\frac{1}{2N}\left(D_{I}N_{J}+D_{J}N_{I}+\eta_{\bar{A}\bar{B}}\partial_{0}(e^{\bar{A}}_{I}e^{\bar{B}}_{j})\right):
\end{equation}
it is easily inferred that bein vectors are evolutionary variables, while connections are not.\\
In 5D, the total group we are dealing with is $SO(4,1)$, which, under suitable assumptions, can be written as
\small
\begin{equation}
SO(4,1)\rightarrow SO(3,1)\otimes SU(2),
\end{equation}
\normalsize
where
\small
\begin{equation}
SO(3,1)\rightarrow SU(2)\otimes SU(2),
\end{equation}
\normalsize
so that
\begin{equation}
SO(4,1)\rightarrow SU(2)\otimes SU(2)\otimes SU(2).
\end{equation}
ADM splittings can be performed on the 5D manifold considered in KK theories \cite{lam2006}. The first splitting gives a $4+1$ reduction, where $1$ refers to the time-like dimension ($x^{0}$), while $4$ to the remaining 4D space-like manifold. On this 4D manifold, a further ADM splitting can be performed, as in ordinary 4D Euclidean space-time: here, the coordinate $x^{5}$ plays a role analogous to that of time in the ordinary theory. It is worth remarking that all the quantities involved in the Gauss-Codacci formula are independent of the fifth coordinate, the Fourier expansion being truncated at the zero-th order, as requested by the cylindrical hypothesis: the residual symmetry of the metric components after the implementation of the KK paradigm is $P^{4}\otimes R$ rather than $P^{4}\otimes U(1)$ \cite{dd1984}; hence,it is possible to perform the second ADM splitting without physically violating the Frobenius theorem.\\ 
In the 4D Euclidean scenario, the part of thr 5D Lorentz group that has to be considered is
\small
\begin{equation}
SO(4) \rightarrow SO(3) \otimes SU(2),
\end{equation}
\normalsize
the degree of freedom concerning $x^{0}$ having been frozen as a result of the first splitting. Two kinds of Ashtekar-like variables can be established\footnote{Lower-case Greek letters refer to 4D space-like world indices, i.e. $\alpha=1,2,3$, while lower-case Latin letters stand for 4D bein indices, i.e. $i=1,2,3$.}
\small
\begin{equation}
C^{i}_{\alpha}=\epsilon^{i}_{jk}\omega^{jk}_{\alpha}
\end{equation}
\begin{equation}
{^{\pm}A}^{i}_{\alpha}={^{\pm}A}^{i5}_{\alpha}=\omega^{i5}_{\alpha}\pm\frac{1}{2}(\beta)C^{i}_{\alpha} 
\end{equation}
\normalsize
where $\epsilon^{ijk}\equiv\epsilon^{i5jk}$, and $(\beta)$ is an Immirzi-like parameter \cite{hols}, but is not strictly needed, since, in this Euclidean metric, it would only render the formalism more complicated. This way, the variables $C^{i}_{\alpha}$ account for the $SO(3)$ group, while he variables ${^{+}A}^{i}_{\alpha}$ for $SU(2)$, thus corroborating the choice $A^{\bar{0}\bar{5}}_{\mu}\equiv0$. Nonetheless, $e^{5}_{\alpha}$ and $e^{i}_{\alpha}$ are evolutionary variables, while ${^{+}A}^{i}_{\alpha}$ and $C^{i}_{\alpha}$ are non-evolutionary variables. In order to avoid this rather unpleasant feature, already envisaged with the analysis of the Gauss-Codacci formula, the two kinds of variables can be mixed up, as, for example,$M^{i}_{\alpha}=C^{i}_{\alpha}+\gamma e^{i}_{\alpha}$ and
$Q^{i}_{\alpha}={^{+}A}^{i}_{\alpha}+\delta \epsilon_{\alpha \nu \rho}e^{i\nu}e^{5\rho}$, where the parameters $\gamma$ and $\beta$ here have the task of mixing the fields up; evolutionary variables would be defined, but the original Ashtekar formalism would be discarded.\\
As a perspective, a more general theory can be looked for, such that the $SU(2)\otimes SU(2)$ group can be built making use of the variables $C^{i}_{\alpha}, e^{i}_{\alpha}, N^{i}, N$, thus recasting an Ashtekar-like formalism, without mixing up ${^{+}A}^{i}_{\alpha}$ and $e^{i}_{\alpha}$.

\end{document}